\documentclass[12pt,draftclsnofoot,onecolumn]{IEEEtran}

\usepackage{amsmath,amssymb,amsmath,amsfonts}
\usepackage{bm}
\usepackage{bbm}
\usepackage{graphicx}
\usepackage{cite}
\usepackage{epstopdf}
\usepackage{balance}
\usepackage{gensymb}
\usepackage{color}
\usepackage{lipsum}
\usepackage{float}
\usepackage{epstopdf}
\usepackage[mathcal]{eucal}
\usepackage{subfiles}
\usepackage[T1]{fontenc}
\usepackage{siunitx}
\usepackage[hyphens]{url}
\usepackage{tabulary}
\usepackage{epsfig,graphics,graphicx,subfig,amssymb,amstext,amsmath,algorithm,algorithmic,wrapfig,multirow}
\usepackage{array}
\usepackage{adjustbox}
\usepackage[dvipsnames]{xcolor}
\usepackage{cite}
\usepackage{times}
\usepackage{placeins}
\usepackage{textcomp}
\usepackage[font=small]{caption}
\usepackage[breaklinks=true]{hyperref}
\usepackage{flushend}
\usepackage{color, colortbl}
\usepackage{tabularx}
\usepackage{bm}
\usepackage{enumerate}
\usepackage{tikz}
\usepackage{pgfplots}
\usepackage{epstopdf}
\usetikzlibrary{shapes,arrows}
\usepackage[utf8]{inputenc}
\usepackage{mathtools}
\usepackage[utf8]{inputenc}
\usepackage{xcolor}
\usepackage{tikz}
\usetikzlibrary{chains,arrows,calc,positioning}
\usepackage{amssymb}
\usepackage{pifont}
\usepackage{soul}
\usepackage{breqn} 
\usepackage{booktabs} 
\usepackage{multirow}
\usepackage{enumitem}
\usepackage{tabulary}
\usepackage[normalem]{ulem}
\usepackage{dblfloatfix}
\usepackage{makecell}
\usepackage{lipsum}
\usepackage{amsthm}
\usepackage{comment}

\usepackage{svg}

\newtheoremstyle{mystyle}
  {}
  {}
  {\itshape}
  {}
  {\bfseries}
  {.}
  { }
  {}
\theoremstyle{mystyle}

\newlength \figwidth
\pgfplotsset{compat=1.16}
\definecolor{bittersweet}{rgb}{1.0, 0.44, 0.37}
\definecolor{glaucous}{rgb}{0.38, 0.51, 0.71}
\definecolor{gainsboro}{rgb}{0.86, 0.86, 0.86}
\definecolor{babyblueeyes}{rgb}{0.63, 0.79, 0.95}
\definecolor{silver}{rgb}{0.75, 0.75, 0.75}
\definecolor{neoncarrot}{rgb}{1.0, 0.64, 0.26}
\definecolor{Gray}{gray}{0.9}
\definecolor{LightCyan}{rgb}{0.88,1,1}
\definecolor{BackgroundLightBlue}{rgb}{0.97,0.97,1}
\definecolor{BackgroundGray}{gray}{0.98}

\usepackage[acronyms,nonumberlist,nopostdot,nomain,nogroupskip]{glossaries}
\newacronym{quic}{QUIC}{Quick UDP Internet Connections}
\newacronym{3gpp}{3GPP}{3rd Generation Partnership Project}
\newacronym{adc}{ADC}{Analog to Digital Converter}
\newacronym{5g}{5G}{5th generation}
\newacronym{aimd}{AIMD}{Additive Increase Multiplicative Decrease}
\newacronym{am}{AM}{Acknowledged Mode}
\newacronym{amc}{AMC}{Adaptive Modulation and Coding}
\newacronym{aqm}{AQM}{Active Queue Management}
\newacronym{awgn}{AGWN}{Additive White Gaussian Noise}
\newacronym{afd}{AFD}{Austin Fire Department}
\newacronym{balia}{BALIA}{Balanced Link Adaptation}
\newacronym{bdp}{BDP}{Bandwidth-Delay Product}
\newacronym{bf}{BF}{Beamforming}
\newacronym{cc}{CC}{Congestion Control}
\newacronym{cdf}{CDF}{Cumulative Distribution Function}
\newacronym{cn}{CN}{Core Network}
\newacronym{cqi}{CQI}{Channel Quality Information}
\newacronym{cp}{CP}{Control Plane}
\newacronym{csirs}{CSI-RS}{Channel State Information - Reference Signal}
\newacronym{dc}{DC}{Dual Connectivity}
\newacronym{dce}{DCE}{Direct Code Execution}
\newacronym{dci}{DCI}{Downlink Control Information}
\newacronym{dl}{DL}{Downlink}
\newacronym{dmr}{DMR}{Deadline Miss Ratio}
\newacronym{dmrs}{DMRS}{DeModulation Reference Signal}
\newacronym{e2e}{E2E}{End-to-End}
\newacronym{ecn}{ECN}{Explicit Congestion Notification}
\newacronym{edf}{EDF}{Earliest Deadline First}
\newacronym{enb}{eNB}{evolved Node Base}
\newacronym{epc}{EPC}{Evolved Packet Core}
\newacronym{es}{ES}{Edge Server}
\newacronym{fdma}{FDMA}{Frequency Division Multiple Access}
\newacronym{fdd}{FDD}{Frequency Division Duplexing}
\newacronym[firstplural=Radio Access Technologies (RATs)]{rat}{RAT}{Radio Access Technology}
\newacronym{fs}{FS}{Fast Switching}
\newacronym{ftp}{FTP}{File Transfer Protocol}
\newacronym{gnb}{gNB}{Next Generation Node Base}
\newacronym{harq}{HARQ}{Hybrid Automatic Repeat reQuest}
\newacronym{hetnet}{HetNet}{Heterogeneous Network}
\newacronym{hh}{HH}{Hard Handover}
\newacronym{hol}{HOL}{Head-of-Line}
\newacronym{ia}{IA}{Initial Access}
\newacronym{imt}{IMT}{International Mobile Telecommunication}
\newacronym{iot}{IoT}{Internet of Things}
\newacronym{los}{LOS}{Line of Sight}
\newacronym{lte}{LTE}{Long Term Evolution}
\newacronym{m2m}{M2M}{Machine to Machine}
\newacronym{mac}{MAC}{Medium Access Control}
\newacronym{mc}{MC}{Multi-Connectivity}
\newacronym{mcs}{MCS}{Modulation and Coding Scheme}
\newacronym{mec}{MEC}{Mobile Edge Cloud}
\newacronym{mi}{MI}{Mutual Information}
\newacronym{mimo}{MIMO}{Multiple Input, Multiple Output}
\newacronym{mmwave}{mmWave}{millimeter wave}
\newacronym{mr}{MR}{Maximum Rate}
\newacronym{mss}{MSS}{Maximum Segment Size}
\newacronym{mtd}{MTD}{Machine-Type Device}
\newacronym{mtu}{MTU}{Maximum Transmission Unit}
\newacronym{nfv}{NFV}{Network Function Virtualization}
\newacronym{nlos}{NLOS}{Non Line of Sight}
\newacronym{nr}{NR}{New Radio}
\newacronym{ofdm}{OFDM}{Orthogonal Frequency Division Multiplexing}
\newacronym{pdcch}{PDCCH}{Physical Downlonk Control Channel}
\newacronym{pdcp}{PDCP}{Packet Data Convergence Protocol}
\newacronym{pdsch}{PDSCH}{Physical Downlink Shared Channel}
\newacronym{pdu}{PDU}{Packet Data Unit}
\newacronym{pf}{PF}{Proportional Fair}
\newacronym{pgw}{PGW}{Packet Gateway}
\newacronym{phy}{PHY}{Physical}
\newacronym{pbch}{PBCH}{Physical Broadcast Channel}
\newacronym[plural=\gls{mme}s,firstplural=Mobility Management Entities (MMEs)]{mme}{MME}{Mobility Management Entity}
\newacronym{prb}{PRB}{Physical Resource Block}
\newacronym{pss}{PSS}{Primary Synchronization Signal}
\newacronym{pucch}{PUCCH}{Physical Uplink Control Channel}
\newacronym{pusch}{PUSCH}{Physical Uplink Shared Channel}
\newacronym{rach}{RACH}{Random Access Channel}
\newacronym{ran}{RAN}{Radio Access Network}
\newacronym{red}{RED}{Robotics Emergency Deployment}
\newacronym{rf}{RF}{Radio Frequency}
\newacronym{rlc}{RLC}{Radio Link Control}
\newacronym{rlf}{RLF}{Radio Link Failure}
\newacronym{rrc}{RRC}{Radio Resource Control}
\newacronym{rrm}{RRM}{Radio Resource Management}
\newacronym{rr}{RR}{Round Robin}
\newacronym{rs}{RS}{Remote Server}
\newacronym{rsrp}{RSRP}{Reference Signal Received Power}
\newacronym{rss}{RSS}{Received Signal Strength}
\newacronym{rtt}{RTT}{Round Trip Time}
\newacronym{rw}{RW}{Receive Window}
\newacronym{rx}{RX}{Receiver}
\newacronym{sa}{SA}{standalone}
\newacronym{sack}{SACK}{Selective Acknowledgment}
\newacronym{sap}{SAP}{Service Access Point}
\newacronym{sch}{SCH}{Secondary Cell Handover}
\newacronym{scoot}{SCOOT}{Split Cycle Offset Optimization Technique}
\newacronym{sdma}{SDMA}{Spatial Division Multiple Access}
\newacronym{sinr}{SINR}{Signal to Interference plus Noise Ratio}
\newacronym{sm}{SM}{Saturation Mode}
\newacronym{snr}{SNR}{Signal to Noise Ratio}
\newacronym{son}{SON}{Self-Organizing Network}
\newacronym{ss}{SS}{Synchronization Signal}
\newacronym{srs}{SRS}{Sounding Reference Signal}
\newacronym{sss}{SSS}{Secondary Synchronization Signal}
\newacronym{tb}{TB}{Transport Block}
\newacronym{tcp}{TCP}{Transmission Control Protocol}
\newacronym{tdd}{TDD}{Time Division Duplexing}
\newacronym{tdma}{TDMA}{Time Division Multiple Access}
\newacronym{tfl}{TfL}{Transport for London}
\newacronym{tm}{TM}{Transparent Mode}
\newacronym{trp}{TRP}{Transmitter Receiver Pair}
\newacronym{tti}{TTI}{Transmission Time Interval}
\newacronym{ttt}{TTT}{Time-to-Trigger}
\newacronym{tx}{TX}{Transmitter}
\newacronym{ue}{UE}{User Equipment}
\newacronym{ul}{UL}{Uplink}
\newacronym{uml}{UML}{Unified Modeling Language}
\newacronym{um}{UM}{Unacknowledged Mode}
\newacronym{utc}{UTC}{Urban Traffic Control}
\newacronym{vm}{VM}{Virtual Machine}
\newacronym{rsrq}{RSRQ}{Reference Signal Received Quality}
\newacronym{rssi}{RSSI}{Received Signal Strength Indicator}
\newacronym{crs}{CRS}{Cell Reference Signal}
\newacronym{comp}{CoMP}{Coordinated Multi-Point}
\newacronym{cran}{C-RAN}{Cloud \acrlong{ran}}
\newacronym{ca}{CA}{Carrier Aggregation}
\newacronym{cco}{CC}{Carrier Component}
\newacronym{nsa}{NSA}{Non Stand Alone}
\newacronym{embb}{eMBB}{Enhanced Mobility Broadband}
\newacronym{bsr}{BSR}{Buffer Status Report}
\newacronym{srb}{SRB}{Service Radio Bearer}
\newacronym{scm}{SCM}{Spatial Channel Model}
\newacronym{sctp}{SCTP}{Stream Control Transmission Protocol}
\newacronym{mptcp}{MPTCP}{Multi-path TCP}
\newacronym{ietf}{IETF}{Internet Engineering Task Force}
\newacronym{os}{OS}{Operating System}
\newacronym{tls}{TLS}{Transport Layer Security}
\newacronym{rfc}{RFC}{Request for Comments}
\newacronym{http}{HTTP}{HyperText Transfer Protocol}
\newacronym{nat}{NAT}{Network Address Translation}
\newacronym{api}{API}{Application Programming Interface}
\newacronym{rto}{RTO}{Retransmission Timeout}
\newacronym{psc}{PSC}{Public Safety Communication}
\newacronym{rpgm}{RPGM}{Reference Point Group Mobility}
\newacronym{ic}{IC}{Incident Command}
\newacronym{rsu}{RSU}{Road Side Unit}
\newacronym{uav}{UAV}{unmanned aerial vehicle}
\newacronym{usv}{USV}{Unmanned Surface Vehicle}
\newacronym{uas}{UAS}{Unmanned Aerial System}
\newacronym{iab}{IAB}{Integrated Access and Backhaul}
\newacronym{qoe}{QoE}{Quality of Experience}
\newacronym{ssim}{SSIM}{Structural Similarity Index}
\newacronym{psnr}{PSNR}{Peak Signal to Noise Ratio}
\newacronym{bs}{BS}{Base Station}
\newacronym{mu}{MU}{Multiple User}
\newacronym{ag}{AG}{Air-to-Ground}
\newacronym{af}{AF}{Array Factor}
\newacronym{ula}{ULA}{Uniform Linear Array}
\newacronym{upa}{UPA}{Uniform Planar Array}
\newacronym{lcs}{LCS}{Local Coordinate System}
\newacronym{psd}{PSD}{Power Spectral Density}
\newacronym{vq}{VQ}{vector quantization}
\newacronym{a2g}{A2G}{air-to-ground}
\newacronym{em}{EM}{electromagnetic}
\newacronym{vae}{VAE}{variational autoencoder}

\makeatletter

 \let\oldforeign@language\foreign@language
 \DeclareRobustCommand{\foreign@language}[1]{%
   \lowercase{\oldforeign@language{#1}}}

\def\nb0{{\mathbf{0}}}
\def\nb1{{\mathbf{1}}}

\makeatother

\IEEEoverridecommandlockouts

\begin{document}
\title{Accordion: A Communication-Aware Machine Learning Framework for Next Generation Networks}

\author{
Fadhel Ayed, Antonio De Domenico, Adrian Garcia-Rodriguez, and David Lopez-Perez\\
 \small{Huawei Technologies, Paris Research Center, 20 quai du Point du Jour, Boulogne Billancourt, France.}
}

\maketitle

\begin{abstract}
In this article, 
we advocate for the design of ad hoc artificial intelligence (AI)/machine learning (ML) models to facilitate their usage in future smart infrastructures based on communication networks. 
To motivate this, 
we first review key operations identified by the 3GPP for transferring AI/ML models through 5G networks and the main existing techniques to reduce their communication overheads.
We also present a novel communication-aware ML framework, 
which we refer to as Accordion, 
that enables an efficient AI/ML model transfer thanks to an overhauled model training and communication protocol. 
We demonstrate the communication-related benefits of Accordion, analyse key performance trade-offs, and discuss potential research directions within this realm.
\end{abstract} 
\section*{Introduction}
\label{sec:introduction}

Artificial intelligence (AI) and machine learning (ML) are becoming ubiquitous in our lives.
At a consumer level, 
most used applications in our smartphones rely on AI/ML algorithms for image recognition and video editing.
Due to their benefits,
vendors and operators are also currently investing on AI/ML models to plan, design, and operate their fifth generation (5G) networks~\cite{ADN}.
From an operational perspective,  
the integration of neural processing units and 
the execution of AI/ML models in smartphones are becoming commonplace to satisfy the latency constraints of complex applications and/or safeguard their data privacy. 
Wireless communications play a central role in this distributed trend, 
\emph{i)} enabling training and/or execution of AI/ML models in {edge} servers, 
and 
\emph{ii)} making them---or their results---available at the user equipment (UE) on-demand~\cite{9357490, 3GPP22874}.

Despite this 
paradigm,
when inspecting popular AI/ML-based applications, 
we can observe that they are increasingly becoming computation-intensive, power-hungry, and memory-demanding~\cite{greenAI}. 
For instance, AlexNet~\cite{krizhevsky2012imagenet} and VGG16~\cite{simonyan2014very} require 724M and 15.5G multiply-accumulate (MAC) operations to classify an image, respectively~\cite{3GPP22874}. 
AlexNet occupies 240 MBytes in memory, 
whereas VGG16 takes up 552 MBytes~\cite{3GPP22874}. 
AI/ML research is also consistent with this development, as it
mainly focuses on enhancing model accuracy,
which in turn requires more processing capabilities~\cite{greenAI}. 
Such trend 
is illustrated by applications where different AI/ML models perform the same task under different conditions with the 
intention to select the most accurate result~\cite{10.1145/3211332.3211336}.

With this landscape, 
the need for downloading and executing large AI/ML models within tight deadlines is translating into stringent capacity and latency requirements on wireless networks, 
and is motivating the design of solutions,
which keep the communication-related needs at bay~\cite{3GPP22874}. 
This need has recently inspired a number of industrial and research developments within both the AI/ML and wireless communities~\cite{9357490}. 
{An important line of work in this direction is the communication efficient distributed (or federated) learning \cite{shi2020communication}, 
where communication constraints mainly impact the training phase of the ML model.}
 
In this paper, 
we first provide a hands-on view on the design of communication-aware AI/ML models. 
We then condense the latest third generation partnership project (3GPP) studies on the transfer of AI/ML models in 5G networks, 
and briefly review the main techniques used to lessen their communication and computational requirements. 
Thereafter,
we introduce Accordion, a communication-aware ML framework designed for efficient transmission of AI/ML models,
whose accuracy can be tailored on-demand, 
and importantly, on-the-run. 
This is achieved by refashioning the AI/ML model training phase so that UEs can execute a fraction of the AI/ML models with adequate performance, 
and incrementally receive additional parts to further enhance accuracy, 
whenever necessary. 
In the remaining of the paper, 
we dive into the implementation details of Accordion,
illustrate its performance when used with a popular AI/ML model, 
and outline ideas meriting further investigation.
\section*{AI/ML Models in 5G}
\label{sec:3GPP}
{The 3GPP has recently investigated  AI/ML model distribution and transfer through 5G networks~\cite{3GPP22874}. 
In this study, the 3GPP recognises
image/speech recognition, media processing/enhancement, automotive systems, robotic control, and computational offloading as the most important use cases.
On the other hand, it has identified
AI/ML model splitting, AI/ML model downloading, and distributed/federated learning as the key operations for the deployment of AI/ML models in a 5G networks.
To show the importance of novel communication-aware AI/ML techniques, 
this section summarizes these key operations for the deployment of AI/ML models in 5G systems.}


\subsection*{AI/ML Model Splitting}
In this technique, 
the execution of an already-trained AI/ML model is divided into two parts: 
A fraction of the model is executed at the UE, 
while the remaining part runs at a network endpoint, 
which then communicates the result to the UE.
\begin{itemize}
\item \emph{The key benefit:} 
Reduction of the computational complexity and power consumption required for the inference process at UEs, 
and/or preservation of the local data privacy. 
Depending on the splitting point, 
the uplink traffic may also be limited with respect to the case where the UE entirely transmits its local data to the network.
\item \emph{Practical application:} 
Consider a case where UE take videos,
which require AI/ML post-processing for enhancements.
Each UE calculates and transmits an intermediate result to the relevant network endpoint. 
The remaining part of the model is then executed at such network endpoint, 
which feeds back the result to the UE.
\end{itemize} 

\subsection*{AI/ML Model Downloading}
UEs 
may not be able to store all the scenario-dependent AI/ML models due to memory constraints. 
Consequently, 
UEs may request a fresh model to the network endpoint to execute a new task and/or operate under new conditions.
As in the previous use case, 
the models are pre-trained at an AI/ML server. 
\begin{itemize}
\item \emph{The key benefit:} 
Availability of a large collection of AI/ML models tailored to a multiplicity of particular tasks and environments,
with the resulting accuracy gain
and alleviation of the memory requirements of UEs.
\item \emph{Practical application:} 
A UE requires identifying a variety of objects from its stored photos, 
which were taken with different background and lighting conditions.
To achieve the highest identification performance, 
such UE requests scenario-specific models to the network endpoint.
\end{itemize}

\subsection*{Distributed/Federated Learning}
In federated learning, 
UEs receive a global model from the network endpoint.
After training with their available local data, 
UEs transmit their updated models back to the network endpoint, 
which then updates the global model via, 
e.g., federated averaging \cite{9141214},
and communicates it back to the UEs. 
\begin{itemize}
\item \emph{The key benefit:} 
Reduction of the communication requirements with respect to the case where all UEs entirely transmit their training data to the network endpoint,
and/or preservation of the UE data privacy at the expense of a computational effort at the UE on local training.
\item \emph{Practical application:} 
When an AI/ML-reliant image recognition application deployed in a large number of UEs reports a poor performance, 
the network endpoint can upgrade the related AI/ML model by combining those newly trained by different UEs via federated learning, 
without accessing their privately stored data.
\end{itemize} 


Overall, 
these use cases make apparent the benefits and need for implementing communication-aware AI/ML techniques,
which reduce the network requirements, 
while keeping performance.
\begin{figure*}[!th]
	\centering
		\includegraphics[width=0.90\textwidth]{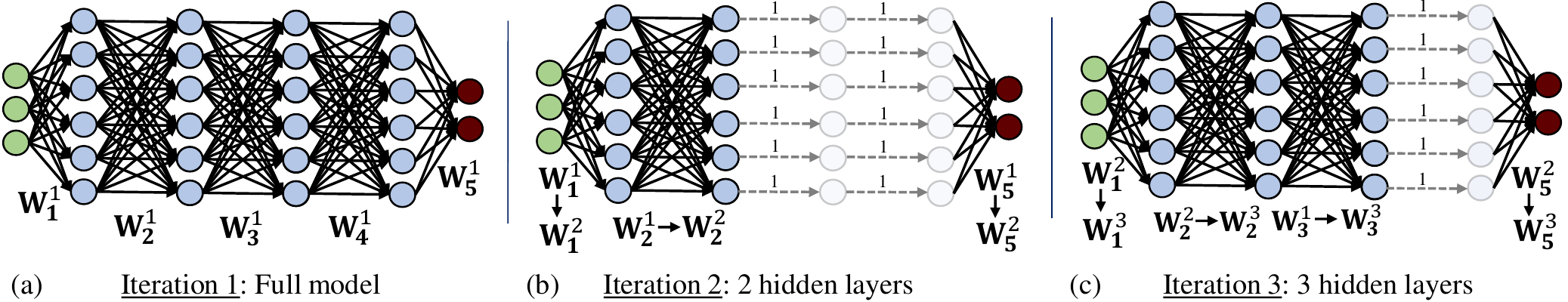}
		\caption{Illustration of the Accordion training. The matrix $\mathbf{W}_{i}^{j}$ describes the model weights of the $i$-th layer at the $j$-th training iteration. (a) In the first training iteration, the weights from all the layers are updated. (b) In the second training iteration, only the weights of two hidden layers and the output layer are updated. (c) In the third training iteration, the weights of three hidden layers and the output layer are updated.}
	\label{fig:trainingCoML}
\end{figure*}
\section*{Reducing the Communication Requirements of AI/ML Models: State-of-the-Art}

The main techniques used to reduce the communication requirements and the computational complexity of AI/ML models are \cite{cheng2017survey, 9043731}:
\begin{itemize}
\item \emph{Quantization} compresses the AI/ML model by reducing the number of bits per model parameter. 
Instead of the typical 32-bit single-precision floating-point format, 
the quantization process usually represents each model parameter with a 16-bit half-precision floating-point format or a 8-bit mini float. 
\item \emph{Pruning} reduces the number of model parameters, 
e.g., by removing weights in a neural network (NN). 
In its simplest implementations, 
pruning can be performed by removing the less important model parameters, 
which usually have small magnitudes. 
\item \emph{Matrix decomposition} relies on modelling the parameters of AI/ML models as matrices, 
e.g., describing the weights connecting two layers of a feed forward neural network (FNN). 
When correlation exists, 
these parameter matrices are low-rank, 
and can be decomposed into smaller matrices with an overall reduction on the number of parameters.
\item \emph{Early Exit of Inference (EEoI)} introduces outputs at different points across a NN. 
During the inference process, 
when an early inference can be performed with high confidence,
EEoI allows obtaining a reliable result without reaching the last layer. 
Note that these methods typically incorporate additional trainable layers~\cite{7979979} or a supplementary network, 
indicating the layers that may be skipped during inference~\cite{8579017}.
{
\item \emph{Knowledge distillation (KD)} uses the information from a large model (teacher) to train a small model (student). 
The student model mimics the teacher model to obtain a competitive or even a superior performance, with a significantly smaller size. 
}
\end{itemize}

Although these techniques can reduce the communication overhead when operating with AI/ML models, 
they share the following drawbacks:
\begin{enumerate}
\item 
Before the UE can execute them,
the AI/ML models need to be either completely downloaded---for quantization, pruning, {KD}, and matrix decomposition---or be comprised of additional trainable parameters with respect to the reference model---for existing EEoI techniques. 
Guaranteeing this within the latency deadline set by the application may not always be feasible due to the large sizes of state-of-the-art AI/ML models and the constraints in the communication link.
\item 
When using quantization, pruning, {KD}, and matrix decomposition,
and AI/ML-driven applications change their performance requirements, 
the UE needs to download a new model with the consequence toll on the communication network.  
Although EEoI can reduce this issue, 
it still suffers from a lack of flexibility,
as its training complexity increases with the number of introduced outputs.
\end{enumerate}

In the following, 
we present a novel communication-aware AI/ML technique addressing these drawbacks.
\section*{Accordion: A Communication-Aware AI/ML Framework}
\label{sec:proposedTechnique}



The cornerstone of Accordion is its new AI/ML model training phase. 
In a nutshell, 
the Accordion training process minimizes the training complexity by considering different configurations of an AI/ML model,
each one comprising a fraction of the full, e.g., NN architecture. 
This process,
illustrated in \figurename~\ref{fig:trainingCoML} for a typical FNN recalls the musical instrument played by compressing or expanding its bellows, 
while pressing buttons or keys.
In more detail,
at each training iteration, 
Accordion only updates the weights of a subset of hidden layers based on a given probability distribution function, 
which can be seen as a model hyper-parameter. 
Although the number of trainable hidden layers per training iteration may be selected uniformly at random in the simplest implementation, 
techniques to perform such selection based on communication- and/or application-related requirements can have significant benefits, 
as described in the following section. 
Importantly, 
the weights of the final layer/s in the original AI/ML architecture are updated in every training iteration, 
and no additional trainable parameters are introduced. 
The non-selected layers are skipped when computing the training error and updating the weights, 
resulting in a reduction of the length of the training phase in average.

It should be noted that the Accordion training process allows the original AI/ML model to be operated with distinct configurations,
each one with a different number of layers, 
and resulting in a different performance and communication overhead---all these,
without requiring a distinct training for each configuration. 
From the proposed methodology,
it follows that the larger the number of hidden layers considered during inference, 
the better the inference accuracy. 

\subsection*{Practical Implementation of Accordion}
\label{sec:practical_implementation}

\begin{figure}[!t]
	\centering
		\includegraphics[width=0.38\textwidth]{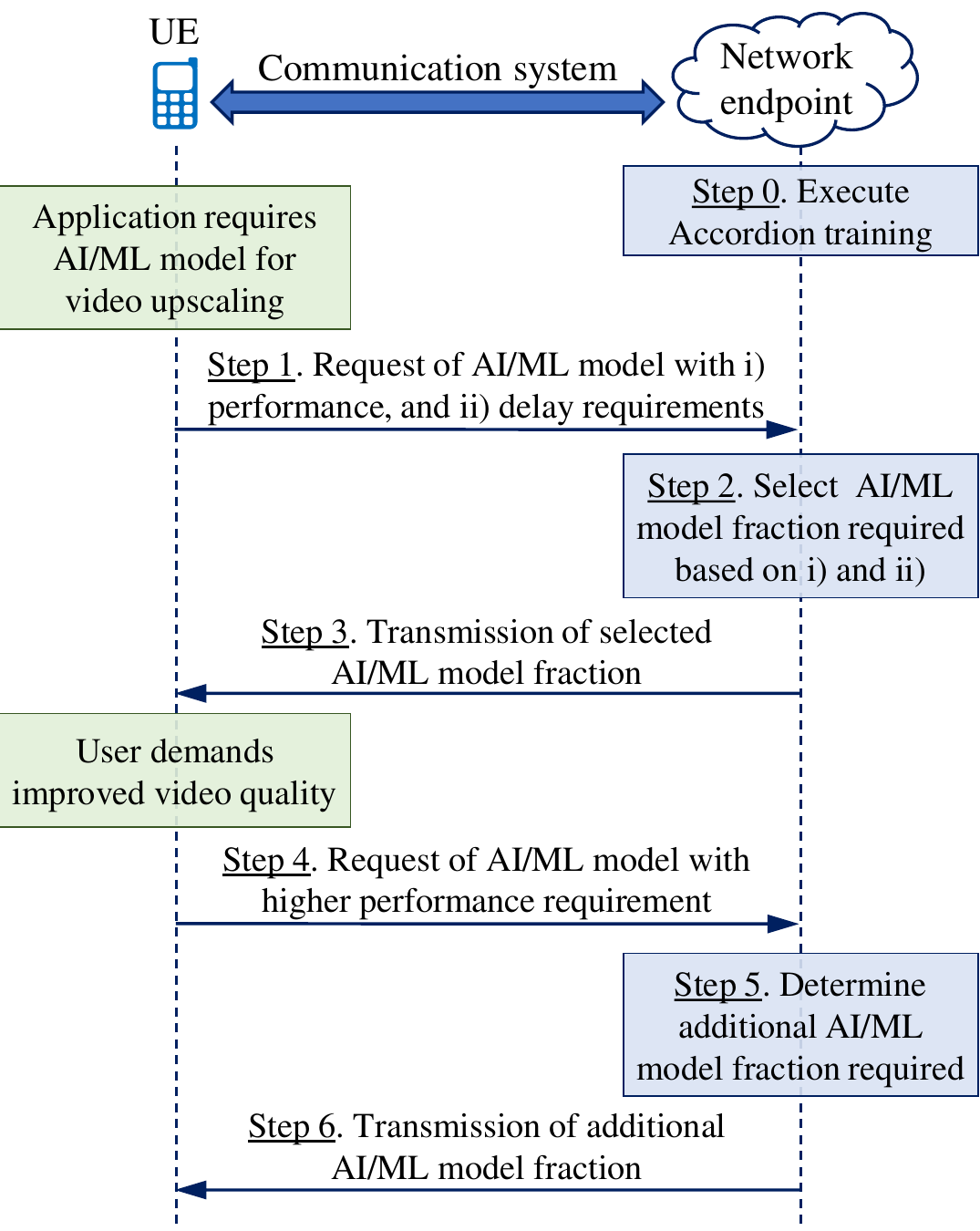}
		\caption{Illustration of the Accordion communication protocol.}
	\label{fig:implementationCoML}
\end{figure}

An implementation example of Accordion is illustrated in \figurename~\ref{fig:implementationCoML}.
{We consider a user attending a concert, 
willing to share a live video on social media. 
Due to poor lighting and sound conditions, 
the UE application needs to run AI/ML algorithms such as media enhancement, face recognition, and song recognition. 
The size of each of these models typically ranges from 3MB to 500MB~\cite{3GPP22874}.} 
In this context, 
{the network edge} server uses Accordion to train in advance the relevant AI/ML model (Step 0), 
where the reduced-size configurations with less layers in the production phase will result in videos with the target resolution but reduced visual quality. 
As mentioned, 
the network endpoint may use a uniform distribution probability to determine the number of layers per training iteration. 
In the following, however, we present two enhanced, but intuitive, communication-aware approaches, 
which exploit a priori available statistical information on the communication network performance and AI/ML-based application requirements:
\begin{enumerate} 
\item \emph{Communication-related criterion:} 
Given a priori knowledge on the distribution of the link throughput{/channel capacity} and the latency tolerated by the application, 
the network endpoint can infer the maximum number of layers of the AI/ML models that can be transmitted within a given deadline. 
The probability of selecting a given number of layers for each training iteration in Accordion can be then made proportional to the frequency with which a given model configuration is expected to be requested, 
based on said network link throughput{/channel capacity} distribution and application latency requirements.  
\item \emph{Performance-related criterion:} 
It is intuitive that the performance of an Accordion-trained AI/ML model depends on its size, 
with larger models offering more accurate results.
Since different UEs may have dissimilar performance requirements, 
the network endpoint can collect those requirements and infer the probability of a UE requiring a specific model size, 
which can also be used during the training phase for designing the policy that governs the number of layers to be updated at each iteration.
\end{enumerate}

As shown in Fig.~\ref{fig:implementationCoML},
after the AI/ML model has been trained, 
{the UE application requests the network endpoint an AI/ML model for video enhancement with size 10MB and a latency requirement of 200 ms (Step 1).}
Thereafter, 
the network endpoint first infers the fraction of the AI/ML model that needs to be distributed to the specific UE by evaluating the model performance and the throughput of the wireless link (Step 2), 
and then transmits it towards the UE (Step 3). 
{Assume that the UE downlink throughput is 240 Mbps, 
then, using Accordion, 
the network endpoint can send only $60\%$ of the model to meet the application accuracy and latency requirements.}

To achieve Step 2, 
the network endpoint evaluated beforehand the performance of Accordion, 
and built a look-up-table,
which maps different model configurations (size and or number of layers) with the resulting model accuracy.
Note that sending the hidden layers after the input and output ones enables the UE to execute a part of the AI/ML model before it is fully received at the cost of a reduced accuracy.
Subsequently, 
the UE communicates with the network endpoint to let it know that it desires to increase the video quality (Step 4). 
Upon reception of this message, 
the network endpoint determines the additional model fraction 
(e.g., the supplementary NN layers) 
required to achieve the requested performance (Step 5), 
and then sends the necessary information to the UE (Step 6). 
In Step 5, 
the same look-up-table as in Step 2 can be reused. 
At a later stage, 
if the terminal is running out of battery or the accuracy requirement decreases, 
the application can decide to operate with a fraction of the overall received model, 
trading off accuracy with computational effort.





\subsection*{The Key Benefits of Accordion}

When compared with the conventional training and model distribution, 
Accordion brings the following key benefits:
\begin{itemize}
\item 
For application-specific accuracy requirements on the AI/ML model,
the network only needs to transmit the fraction of the model required to meet such requirements.
\item 
When applications require an enhanced accuracy of the AI/ML model,
they can ask for it, ad-hoc,
and the network endpoint share the already-computed, additional parts of the AI/ML model,  
without the need to train and share a new model or additional model parameters. 
\item 
Applications with stringent latency constraints can execute a fraction of the model prior to its complete reception whenever necessary, 
at the cost of reduced performance.
\item 
Even when having access to the complete AI/ML model, 
devices with limited e.g., battery capabilities can still execute a fraction of the AI/ML model.
\item 
The computational complexity of the AI/ML model training is reduced, 
as Accordion requires no additional trainable parameters,
and in each training iteration, 
only a subset of hidden layers is trained.
\item 
Standard approaches to reduce communication and computational requirements of AI/ML models can be implemented together with Accordion in a transparent manner.
\end{itemize}

\section*{Performance Evaluation of Accordion}

\label{sec:perfEvaluation}
\begin{figure*}[!t]
	\centering
		\includegraphics[width=0.98\textwidth]{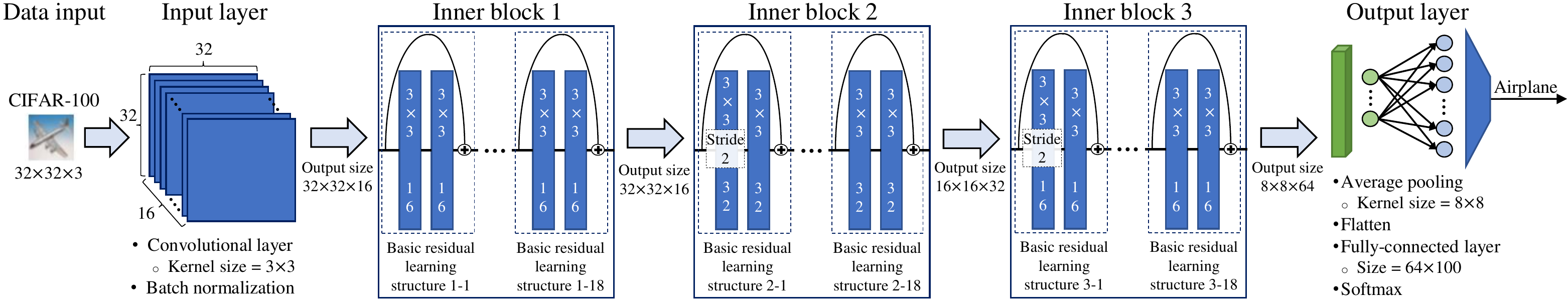}
		\caption{Illustration of the ResNet-110 architecture for the CIFAR-100 dataset \cite{He_2016_CVPR}.}
	\label{fig:ResNet}
\end{figure*}

In this section, 
we illustrate the performance of Accordion over the standard ResNet-110 (resp. ResNet-56) architecture when classifying the CIFAR-100 (resp. CIFAR-10) dataset.
We chose these architectures to demonstrate that the Accordion principle can be readily applied to modern AI/ML frameworks.
The CIFAR-100 dataset includes 100 classes and 600 $32 \times 32$ color images per class. 
The CIFAR-10 dataset includes 10 classes and 6000 $32 \times 32$ color images per class. 
The ResNet 110 (resp. ResNet 56) architecture is comprised of 110 layers (resp. 56 layers) decomposed in basic residual learning structures (BRLSs): 
1 input layer, 1 output layer, and 3 inner blocks of 18 BRLSs (resp. 9 BRLSs), 
each with 2 consecutive convolutional layers and an identity shortcut. 
The architecture of ResNet-110 is illustrated in \figurename~\ref{fig:ResNet} \cite{He_2016_CVPR}. 
Overall, ResNet 110 (resp. ResNet 56) has a total of 1.7 million (resp. 862 thousand) parameters and a size of 111 Mbits (resp. 55 Mbits), 
and requires 255 millions (resp. 127 millions) of MAC operations.

Using these architectures, 
we compare the following 3 schemes,
which trade off classification accuracy for model size to deal with the communication network constraints and AI/ML application requirements: 
baseline ResNet, CoML, and BlockCoML. 
The training phase in ResNet always involves all the BRLSs, 
while in CoML and BlockCoML, 
at each training iteration, 
the number of trained BRLSs follows a random distribution, 
which governs how the performance of these schemes is impacted by the fraction of the model available at the UE. 
Specifically, in our experiments, 
we consider two simple distributions for training CoML and BlockCoML.
In the first one, 
referred to as CoML-05 and BlockCoML-05,  
the probability of training the full model is $50\%$,
while in the second one,
referred to as CoML-03 and BlockCoML-03, 
this probability is $30\%$.
In both cases, 
when the full model is not trained, 
the number of BRLSs to be trained is selected according to a uniform distribution. 

More details on the training and transmission of the AI/ML models based on these schemes are provided in the following:
\begin{itemize}
\item \emph{ResNet:} 
At each training iteration 
all the BRLSs are updated to minimize the classification error of the complete model. 
When the network endpoint wants to transmit a model of reduced size to satisfy application requirements and/or network constraints,
it removes BLRSs from the inner blocks---from right to left---,
i.e., ignoring first the layers of the third inner-block, then those of the second block, and finally those of the first block.
\item \emph{CoML:} 
Trained to minimize the average classification error over different configurations by skipping BRLSs from right to left.
At each training iteration, 
based on the number of selected BRLSs,
CoML first skips the necessary BRLSs of the third inner-block, 
and then those of the second and first blocks, 
if needed be. 
The network endpoint follows the same approach when aiming to transmit a fraction of the complete model to the UE, 
i.e., skipping BLRSs starting from the third inner-block.
For instance, if only 66$\%$ of the full model has to be sent, 
the full third inner-block is not transmitted to the UE.
\item \emph{BlockCoML:} 
Trained to minimize the average classification error over different configurations by skipping a subset of layers in each inner block.
This means that, 
at each training iteration, 
the same number of BRLSs is trained in each inner-block.
The same approach is followed when a fraction of the complete model is transmitted to the UE. 
For instance, 
when wishing to transmit 66$\%$ of the full model, 
12 and 6 BRLSs for each of the three inner blocks are transmitted in ResNet 110 and ResNet 56, respectively.
\end{itemize}

As shown in Fig. \ref{fig:ResNet}, 
the layers of the third inner block have four times more parameters than those of the second,
which in turn, 
has four times more parameters than those of the first inner block. 
Therefore, by skipping first the layers of the last inner block, 
CoML favors smaller models than BlockCoML. 
Specifically, we can order the models according to their size, 
from the smallest to the largest, 
as follows: 
CoML-03, CoML-05, BlockCoML-03, and BlockCoML-05. 
However, this is not true in terms of computational complexity,
since all layers require the same number of MAC operations.

It is well known that the tuning of the hyper-parameters plays a pivotal role in modern AI/ML. 
It is as crucial as the choice of architecture,
and the most time-consuming task of the NN training phase. 
To highlight the practicality of the Accordion training strategy, 
we use the standard hyper-parameters of ResNets,
showing that the proposed training procedure can be used without requiring additional hyper-parameters optimization. 
Specifically, 
we train the parameters with stochastic gradient descent with 164 epochs, 
a mini-batch size of 124 images, 
and an initial learning rate of 0.1, 
which is divided by 10 at iteration 82 and then again at iteration 123. { 
At the last epoch, 
we have observed that all the tested models have converged, 
indicating that Accordion does not negatively affects the convergence rate. 
Not only that, since Accordion requires to update less parameters at each training iteration, 
it takes less time to converge than the baseline solutions. 
Specifically, 
the gain in training time is inversely proportional to the average number of layers trained in each iteration.}

\subsection*{Performance Results}

\begin{figure}[!t]
	\centering
	\subfloat[][CIFAR-10]{
	    \includegraphics[width=0.48\textwidth]{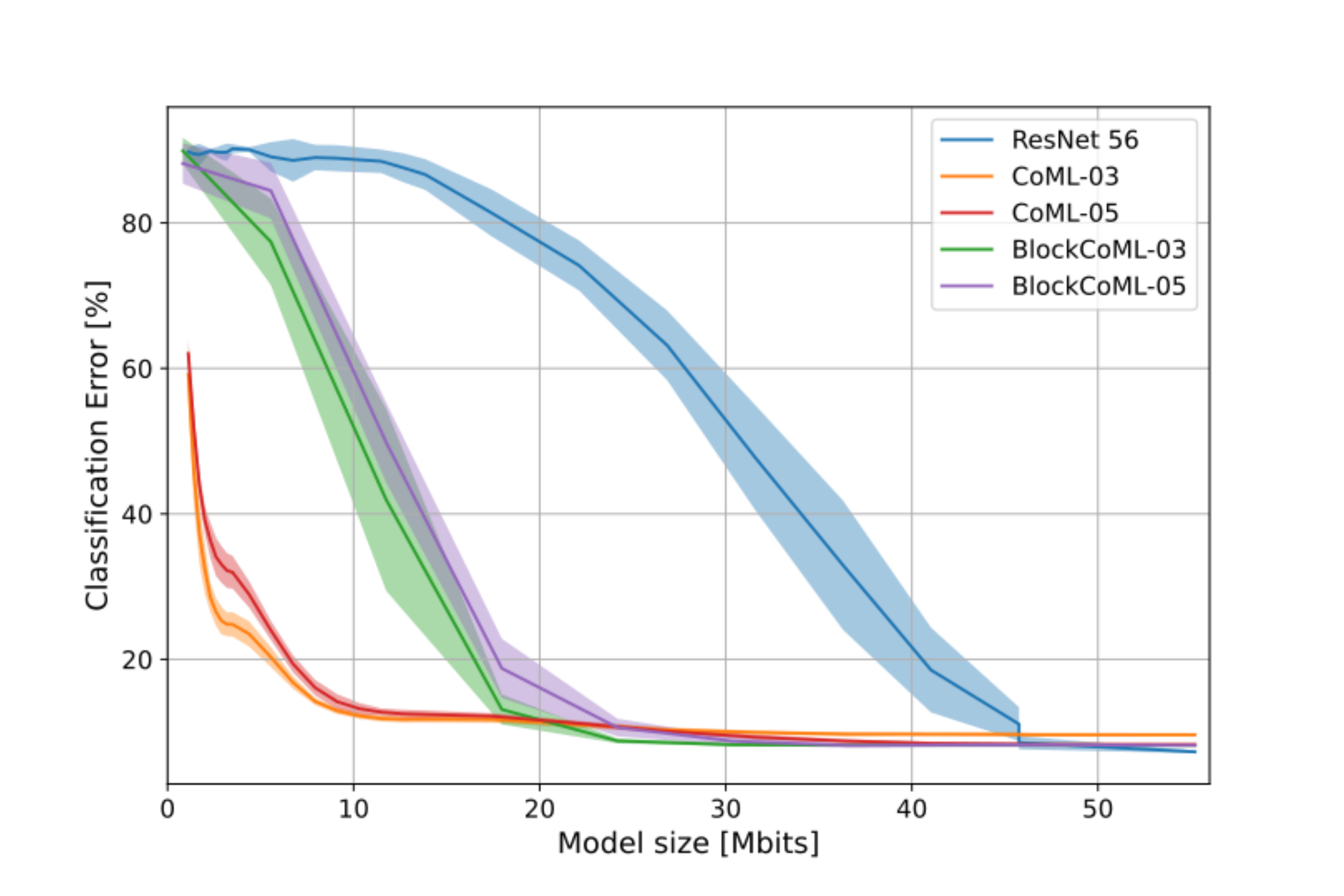}}
	    \hfill
    \subfloat[][CIFAR-100]{
	    \includegraphics[width=0.48\textwidth]{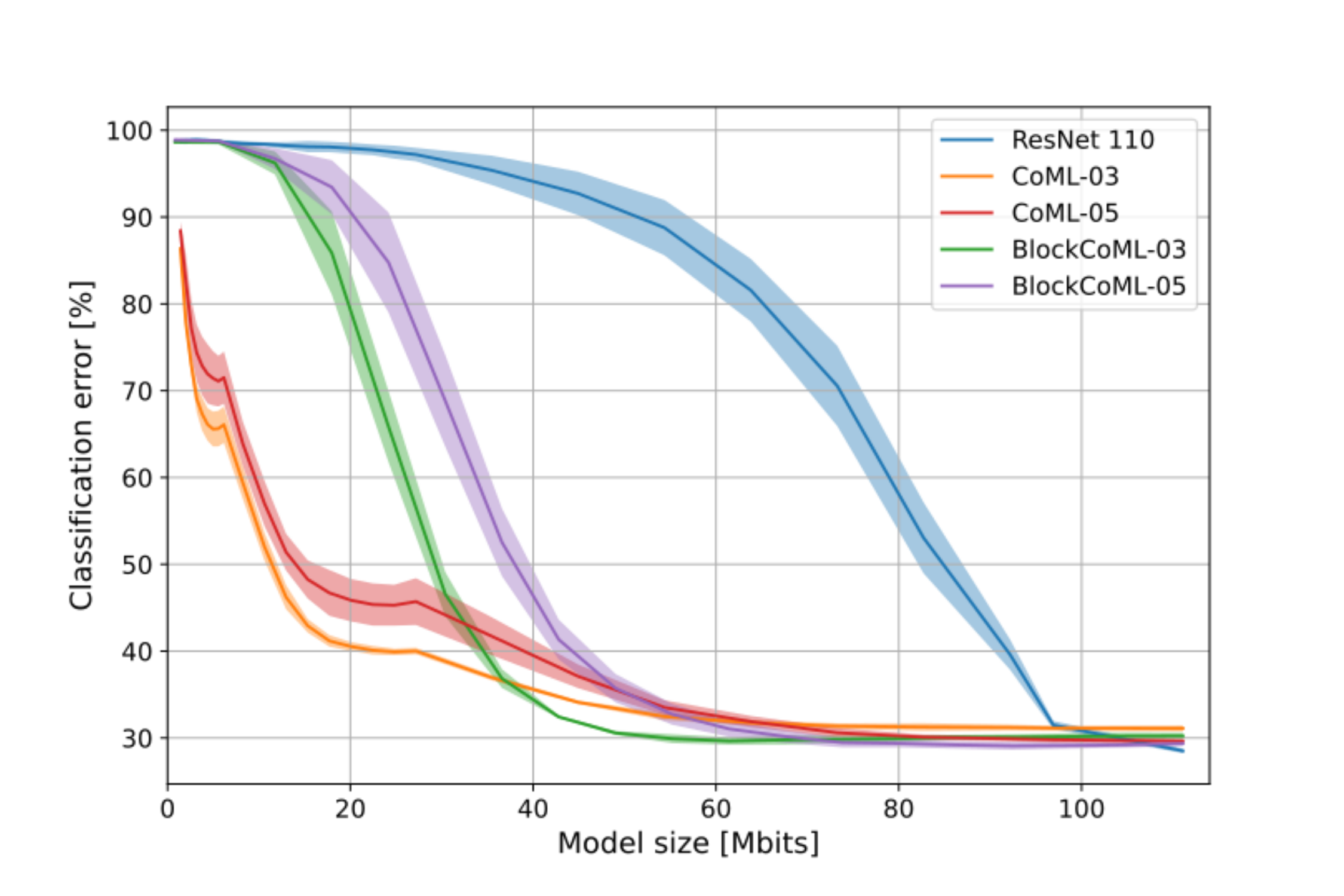}}
	\caption{Classification error as a function of the portion of the model available (in Mbits) using CoML, BlockCoML, and ResNet for the datasets a) CIFAR-10 and b) CIFAR-100. The solid lines represent the mean over five runs, the shaded area illustrates the performance standard deviation.}
	\label{fig:ResNet_perf}
\end{figure}

\figurename~\ref{fig:ResNet_perf} illustrates the classification error of the different schemes over CIFAR-10 and CIFAR-100 datasets as a function of the size of the model transmitted to the UE. 
For each of them, 
we show the mean and standard deviation of the error computed using 5 independent runs. 
Note that the sizes and the architectures of the complete models for each of the schemes are the same,
as they coincide with the ones of ResNet 110 (111 Mbits) and ResNet 56 (55 Mbits),
but that their performances differ, 
due to the specific training processes described in the previous section.  
Although ResNet has been trained to maximize the performance of the complete model, 
CoML and BlockCoML are trained to perform even when their respective full models are not available at the UE. Accordingly, 
we can see that, 
in CIFAR-10 and CIFAR-100, 
the classification errors of ResNet 110 and ResNet 56 quickly increase when the full models are not available at the UE, 
e.g., with 50$\%$ of the full model, 
ResNet 110 and ResNet 56 achieve 88$\%$ and 63$\%$ of error with a loss of 208$\%$ and 89$\%$, respectively.
Importantly, 
we can observe that, 
when only a small fraction of the models is available at the UE, 
the empirical results validate our intuition, 
and the approaches using Accordion significantly outperform the state of the art.
As example, 
to reach a classification error rate of $50\%$ on CIFAR-100, 
and with respect to a scheme requiring to share the complete model,
ResNet 110, BlockCoML-05, BlockCoML-03, CoML-05, and CoML-03 can afford a $25\%$, $66\%$, $75\%$, $89\%$, and $90\%$ network throughput reduction for model sharing, respectively,
due to the reduced model size. 
In terms of the MAC operations required at the UE for the inference,
such models lead to a computational complexity reduction of $11.5\%$, $69.2\%$, $73.1\%$, $57.7\%$, and $61.5\%$,  respectively. 
Note that, 
due to the way in which layers are skipped during the training, 
BlockCoML outperforms CoML in terms of computational complexity reduction, 
while CoML outperforms BlockCoML in terms of throughput reduction. 
{ 
Accordingly, during lightly loaded hours, 
when network capacity is larger than the traffic load, 
the network can reduce the overall computational complexity training ML models using BlockCoML. 
In contrast, during the peak traffic hours, 
CoML is preferable.}

As an example, 
if we consider a 5G link with a throughput of 100 Mbps and an AI/ML-based application that requires a model trained on CIFAR-100 to be download within 0.3 s, 
ResNet 110, BlockCoML-05, BlockCoML-03, CoML-05, and CoML-03 would reach a classification error of $96\%$, $69\%$, $46\%$, $43\%$, $38\%$ respectively.
This clearly shows the shortcomings of ResNet 110 and benefits of the proposed schemes. 
One can argue that these trends change when we increase the available throughput or relax the latency constraint, 
thus increasing the available fraction of the models at the UE. 
However, notice that the loss of the proposed schemes with respect to the standard ResNet with the full model is relatively limited in this case,  
as the error increase ranges from $3\%$ to $9\%$ when using BlockCoML-05 and CoML-03, respectively.

 {
It is worth highlighting that when accuracy requirements and communication constraints are static and known in advance, 
standard approaches for reducing the model size are more effective than Accordion. 
For example, 
using quantization, 
and compressing the model parameters from 32bits to 8bits, 
i.e., reducing the model size by $75\%$, 
ResNet achieves an error rate of $8\%$ on Cifar10. 
Similarly, when pruning $75\%$ of the weights with the smallest absolute value in the original model, 
and then further training the remaining $25\%$ of the weights, 
ResNet achieves an error rate of $10\%$. 
For comparison, 
CoML-30 and CoML-50 result in approximately $12\%$ error with $75\%$ model size reduction. 
Nevertheless, as previously mentioned, 
the main advantage of our proposed framework lays in its enhanced flexibility with respect to these classic mechanisms, 
as it enables the system to efficiently adapt ---in real time--- the transmitted layers to the UE's requirements without the need to send a new model from scratch,
as it would be required by the mentioned state of the art. 
Besides, we conclude highlighting that, 
in principle, 
Accordion is fully compatible with other model size reduction methods to further reduce the communication overhead.}

\section*{Future Works on Communication-aware AI/ML Models}


In this section, 
we highlight further research to pave the way to communication-aware AI/ML techniques.

\subsection*{Accordion in Deep Learning Architectures with Dimensionality Mismatches}

A key feature of Accordion is that the weights of the final layer/s present in the original AI/ML architecture are trained at every iteration. 
In a large variety of architectures such as ResNets, VGGs, AlexNets, and DenseNets, 
Accordion can be directly implemented by just removing one of the many intermediate layers with the same number of inputs and outputs, 
as shown in the previous section. 
{When integrating Accordion in ResNets, 
we use average pooling for down-sampling before each inner block, 
which does not require additional trainable parameters.}
In other architectures where removing intermediate layers necessarily leads to dimensionality mismatches, 
Accordion may still be applied by introducing additional---trainable or non-trainable---linear layers.

\subsection*{Generalization to AI/ML Models}

Although the training part of the Accordion framework focuses on deep learning models due to their pervasiveness, 
its communication protocol (see Fig. \ref{fig:implementationCoML}) is general,
and may be adapted to a variety of AI/ML models.
A prime example is that of ensemble AI/ML methods based on bagging, 
like random forests or boosting, 
such as adaptive or gradient boosting techniques \cite{gonzalez2020practical}. 
These methods are particularly efficient from a communication perspective, 
as they consist of multiple learners that combine their outputs to provide better performance than each of the individuals. 
Since fully trained ensemble learning methods can be executed with a fraction of the learners, 
and their performance generally grows with the number of available learners, 
their implementation is Accordion friendly, 
and may be preferred over NNs whenever feasible.
{In contrast, 
more complex neural network architectures like attention based models, 
e.g. transformers, 
require more careful investigation to integrate Accordion principles.}

\subsection*{Derivation of Accordion policies with performance guarantee}

In this manuscript, 
we have presented solutions to derive distribution probabilities,
which determine at each training iteration of Accordion the layers to be updated. 
However, in future works, 
it is necessary to design policies with performance guarantee, 
in terms of convergence time, accuracy, and communication requirements. 
{In addition, 
it is necessary to understand the end to end energy efficiency for each of these policies and related baseline solutions, 
considering the cost of training, communication, and inference phases.}
The availability of use-case specific policies with proved performance will facilitate the deployment of Accordion in future intelligent systems.

{\subsection*{Accordion in Federated Learning}
We believe that Accordion can provide significant advantages in the the federated learning setting. 
Consider the case where multiple UEs with local data have different computational and communication capabilities. 
With Accordion, 
each UE could train a different (fixed) model configuration tailored to its capabilities, 
e.g., a different number of layers. 
As a result, 
agents with limited capabilities would need to communicate and update only a small fraction of the layers. However, new challenges arise to achieve this goal with respect to the baseline Accordion logic presented in the paper, 
where training is realized at the network side. 
More specifically, 
open research challenges are related to the impact of the wireless channel on the training phase, the power consumption at the end users due to the distributed model training, and the need for accurate aggregation of data from different sources, potentially in an asynchronous manner.}
\section*{Conclusions}
The integration of AI/ML models and wireless networks is progressing at a rapid pace. In this manuscript, we have introduced a novel communication-aware ML framework, which we refer to as Accordion, tailored to enable efficient distribution of NN models in a network. Specifically, thanks to an innovative training, in contrast to standard ML models, Accordion supports stringent application latency constraints and adequate accuracy performance by allowing UEs to execute a fraction of the original AI/ML model. In addition, it enables to improve the application performance by transferring additional parts of the AI/ML model, online, when required. 
\section*{Biographies}
\begin{IEEEbiographynophoto}
    {Fadhel Ayed}
is a Senior Researcher with Huawei Technologies, France. His research interests include resource efficient machine learning, stochastic processes, and mathematical foundations of deep learning. 
\end{IEEEbiographynophoto}

\begin{IEEEbiographynophoto}
    {Antonio De Domenico}
is a Senior Researcher with Huawei Technologies, France. His research interests include heterogeneous wireless networks, machine learning, and green communications. 
\end{IEEEbiographynophoto}

\begin{IEEEbiographynophoto}
{Adrian Garcia-Rodriguez} is a Senior Researcher with Huawei Technologies, France. His research interests include heterogeneous wireless networks, machine learning, and UAV communications.
\end{IEEEbiographynophoto}

\begin{IEEEbiographynophoto}
{David Lopez-Perez} is an Expert and a Technical
Lead with Huawei Technologies, France. His interests are on both cellular and Wi-Fi networks, network performance analysis, network planning and optimization, as well as technology and feature development.
\end{IEEEbiographynophoto}


\bibliographystyle{IEEEtran}

\bibliography{main}

\end{document}